 \documentclass[onecollarge,natbib]{svjour2}

 \usepackage{graphicx}
\usepackage{epsfig}

\def\sla#1{\rlap\slash #1}

\newcommand{\be}{\begin{equation}}
\newcommand{\ee}{\end{equation}}
\newcommand{\cc}{\c c}

\newcommand{\bee}{\begin{eqnarray}}
\newcommand{\eee}{\end{eqnarray}}

\bibpunct{[}{]}{;}{n}{}{,} 
\smartqed  
\usepackage{graphicx,amsmath,amssymb}
\usepackage{color}
\usepackage{enumerate,array}

\usepackage{multicol}

\definecolor{navyblue}{rgb}{0.3,0.3,1}
\definecolor{purple}{rgb}{0.6,0,0.5}
\usepackage[colorlinks=true, pdfstartview=FitV, 
linkcolor=purple, citecolor= navyblue, urlcolor=navyblue]{hyperref}

\journalname{Few Body Systems}

\begin{document}

\title{Light-Front Spin-1 Model: Parameters Dependence\footnote{Presented by Clayton Santos Mello at LIGHT-CONE 2014,~May,~NCSU-USA}}  
\author{Clayton S. Mello \and Anac\'e N. da Silva \and J.~P.~B.~C.~de~Melo 
\and T. Frederico}
\institute{J. P. B. C. de Melo and Anac\'e N. Silva \at
                Laboratorio de F\'isica 
Te\'orica e Computa\cc\~ao Cient\'ifica, 
Universidade Cruzeiro do Sul, 01506-000, S\~ao Paulo, Brazil. \\
                \and  Clayton S. Mello  and T. Frederico \at 
Instituto Tecnol\'ogico de Aeron\'autica, DCTA, 
12228-900, S\~ao Jos\'e dos Campos, Brazil.}
\date{Version of \today}
\maketitle

\begin{abstract}
We study the structure of the  $\rho$-meson within a light-front model with constituent quark degrees of freedom.  
We calculate electroweak static observables: magnetic and 
quadrupole moments, decay constant and charge  radius. The prescription used to 
compute the electroweak quantities is free of zero modes,  which 
makes the calculation implicitly covariant. We compare the results of our model with other ones  found in the literature. 
Our model parameters give a decay constant close to the experimental one.
\keywords{$\rho-$meson \and Light-Front  model \and Electromagnetic structure 
\and Decay constant }
\end{abstract}

\section{Introduction}
\label{intro}

A central question today in physics is to understand the subatomic structure of matter in
terms of fundamental degrees of freedom, i.e, quarks and gluons~\cite{PDG2014}. 
The standard model for the strong interaction, namely, quantum chromodynamics~(QCD), 
with quarks and gluons~\cite{Yndurain,Smilga}, exhibit confinement, beyond a perturbative expansion. Calculations with Lattice QCD and 
Schwinger-Dyson approaches  are performed in Euclidian space, while  it is still a challenge to extract information of the hadron in 
Minkowski space. In this sense, it is still useful to describe the composite states in constituent quark models~(CQM),
which are defined in Minkowski space, and driven towards  experiments  (see e.g.~\cite{Close1980}). 

A vast literature is devoted to investigate the hadronic electromagnetic 
structure ( see e.g. \cite{Roberts96,Pacheco97,deMelo99,Choi99,
deMelo2003,Pacheco2002,deMelo2004,Otoniel2012}) since it provides a laboratory to deeper our understanding  of QCD. 
A natural framework to combine the quantum field theory~(QFT) and CQM to describe bound states is the 
 light-front quantization~(LFQCM)~\cite{Brodsky98,Harindranath2000,Beane2013}. 
One can resort to the usual covariant formulation of  quantum field theory and the light-front 
 quantisation  to test the  nonperturbative properties of QCD. 
 
The aim of this contribution is to study the sensitivity of static electroweak properties
of  the $\rho$-meson  with the model parameters of an ansatz of the  covariant Bethe-Salpeter
 amplitude with constituent quarks proposed in Ref. \cite{Pacheco97}. The  variables used  
 to evaluate numerically the loop integrals is the light-front momentum, and the integrals 
 are reduced to three-dimensional ones, by analytical integration 
 on the minus component of the loop momentum, relating the calculations 
 of the electroweak observables with the present model
  performed in a light-front framework considers only  valence contributions.  In the case of the electromagnetic observables, we use a 
  prescription for computing 
the elastic  form factors which was shown to be free of non-valence  contributions in the Drell-Yan frame (see e.g. \cite{Pacheco2012}). We also 
found that the vector meson decay constant in the present model is free of a zero-mode when computed with the plus component of the current operator.
 In our systematic study we address 
 the dependence of the static electromagnetic    observables and decay constant with the two model parameters, namely the 
 constituent quark and the regulator masses.  In addition, we compare our results with the ones obtained with different  models.

\section{Light-Front model, electromagnetic form factors and decay constant } 

The general framework for the light-front constituent quark model adopted here has been used to study
several properties of hadronic states, like in the case of mesons and baryons in the vacuum~\cite{Jaus1990,Pacheco2009} 
or in nuclear matter \cite{Kazuo2014}. Electromagnetic form factors of composite vector particles within the light-front framework 
 have been addressed in many works 
\cite{Frankfurt79,Inna84,Chung88,Inna89,Brodsky92,Frankfurt93,Cardarelli95,Ji2001,Bakker2001}, 
where a
specific frame with momentum transfer $q^+=q^0+q^3=0$ (Drell-Yan condition) 
is chosen to compute the matrix elements of  the plus component of the current operator. 
Here, we will use this frame and current 
 component, with the choice of matrix elements suggested in
 \cite{Inna89}, which has been shown to be free of zero modes \cite{Melo2004,Choi2004,Pacheco2012}. 
 In this case, the   computation is performed 
 only in the valence region. We will return to this point later in this section.

The $\rho-q\bar{q}$ vertex model for an on-mass-shell meson, is 
the same one proposed  in reference~\cite{Pacheco97}:  
\begin{eqnarray}
\Gamma^\mu = \left[ \gamma^\mu - 
\frac{m_{\rho}}{2}~
\frac{2 k^{\mu} -P^{\mu}}{(P^\mu k_\mu + m\, m_\rho) -\imath \epsilon }     
\right]~,
 \end{eqnarray}
where the quark momentum is  $k^{\mu}$, the constituent quark mass is $m$, 
and the $\rho$ meson mass is $m_\rho.$  
 
 The electromagnetic form factors 
 of a spin-1 particle are computed from a linear combination of the
 matrix elements of the current, $J^{\mu}_{ji}$ with $i$ and $j$ indexing the polarization states. The 
 general covariant expression of  the current is~\cite{Frankfurt79}:
 \begin{equation}
 J_{\alpha \beta}^{\mu}=\left[F_1(q^2)g_{\alpha \beta} -F_2(q^2)
 \frac{q_{\alpha}q_{\beta}}{2 m_{\rho}^2}\right] (p^\mu + p^{\prime \mu})
  - F_3(q^2)
 (q_\alpha g_\beta^\mu- q_\beta g_\alpha^\mu) \ ,  
\label{eq:curr1}
 \end{equation}
 where $m_\rho$ is the rho meson mass, $q^\mu$ is the momentum transfer, 
 and $P^\mu$ is the sum of the initial and final momentum. In the impulse approximation, the plus component of the 
 electromagnetic current,~${ J}^+_{ji}$, is:
\begin{equation}
J^+_{ji} =  \imath  \int\frac{d^4k}{(2\pi)^4}
 \frac{ Tr[\epsilon^{'\nu}_j \Gamma_{\beta}(k,k-p_f)
(\sla{k}-\sla{p_f} +m)  
\gamma^{+} 
(\sla{k}-\sla{p_i}+m) \epsilon^\mu_i \Gamma_{\alpha}(k,k-p_i)
(\sla{k}+m)]
\Lambda(k,p_f)\Lambda(k,p_i) }
{(k^2 - m^2+\imath \epsilon)  
((k-p_i)^2 - m^2+\imath\epsilon) 
((k-p_f)^2 - m^2+\imath \epsilon )}~,
\label{current}
\end{equation}
where ${\epsilon^\prime_j}$ and ${\epsilon_i}$ are the
polarization four-vectors of the final and initial states,
respectively.  In the cartesian instant form spin basis the initial polarization state is given by: 
\begin{equation}
 \epsilon^\mu_x=(-\sqrt{\eta},\sqrt{1+\eta},0,0),~\epsilon^\mu_y=(0,0,1,0),
 ~\epsilon^\mu_z=(0,0,0,1)~, \label{polar}
\end{equation}
and for the final state is given by:
\begin{equation}
 \epsilon^\mu_x=(\sqrt{\eta},\sqrt{1+\eta},0,0),~\epsilon^{\prime\mu}_y=(0,0,1,0),
 ~\epsilon^{\prime\mu}_z=(0,0,0,1)~,
\end{equation}
where $\eta=-q^2/4m^2_\rho$. 
The function~$\Lambda(k,p)=N/((p-k)^2-m^2_R+\imath \epsilon)^2$ 
 reguralizes the loop integral in~Eq.(\ref{current}). The regulator mass is $m_R$ 
 and the normalisation factor  $N$ is fixed by the condition of unit charge.

The Breit-frame is used in the numerical calculations, with the choice $p^\mu=(p^0,-q_x/2,0,0)$ for the initial state and 
$p^{\prime\mu}=(p^0,q_x/2,0,0)$ for the final state, such that the momentum transfer $q^\mu=(p^{\prime^\mu}-p^\mu)$ satisfies the
the Drell-Yan condition,~$q^+=0$. In this case four matrix elements  are independent and 
 only three electromagnetic  form factors exists. The physical constraint is expressed by the angular condition, which in the light-front spin basis is written as:
  \begin{equation}
\Delta(q^2)=(1+2 \eta) I^{+}_{11}+I^{+}_{1-1} - 
\sqrt{8 \eta} I^{+}_{10} -
I^{+}_{00} \ = \ (1 + \eta)(J^+_{yy}-J^+_{zz})=0  \ .
\label{eq:ang}
\end{equation} 
This relation is violated in the present model by the presence of zero-modes not accounted by the naive integration in the loop momentum $k^-$, 
which leads only  to the valence region contribution (see e.g. \cite{Pacheco2012}). In principle, if the angular condition is satisfied, 
  we have the freedom  to extract in different ways the form factors from the matrix elements of the  current 
~\cite{Inna84,Chung88,Brodsky92,Frankfurt93}. 
 
 In the present work we calculate the electromagnetic observables,
 ~$G_0$,~$G_1$ and $G_2$, without  ambiguities due to zero-mode contributions 
 by using the prescription proposed in \cite{Inna84,Inna89}, which can be also computed in the 
instant form basis,  as well as in the light-front spin basis:
\begin{eqnarray}
&&G_0^{GK} = \frac{1}{3}[(3-2 \eta) I^{+}_{11}+ 2 \sqrt{2 \eta} I^{+}_{10} 
+  I^{+}_{1-1}] 
=  \frac{1}{3}[J_{xx}^{+} +(2 -\eta) J_{yy}^{+} 
+ \eta  J_{zz}^{+}], \nonumber \\
&&G_1^{GK} =  2 [I^{+}_{11}-\frac{1}{ \sqrt{2 \eta}} I^{+}_{10}]
=J_{yy}^{+} -  J_{zz}^{+} - \frac{J_{zx}^{+}}{\sqrt{\eta}},
\nonumber \\ 
&&G_2^{GK}=\frac{2 \sqrt{2}}{3}[- \eta I^{+}_{11}+
\sqrt{2 \eta} I^{+}_{10} -  I^{+}_{1-1}] 
=  \frac{\sqrt{2}}{3}[J_{xx}^{+}-(1+\eta) J_{yy}^{+} 
+ \eta  J_{zz}^{+}] \ .
\label{inna}
\end{eqnarray}
The prescription \cite{Inna84,Inna89} amounts to eliminate the matrix element $I_{00}^+$ in the computation of the form factors.
In \cite{Melo2004,Choi2004,Pacheco2012}, it was demonstrated that the zero-modes or non-valence contributions are present only  in  $I^+_{00}$, and
therefore the combinations expressed in Eq. (\ref{inna}) can be computed relying only on the valence region.

The low-energy electromagnetic observables  considered here are
the  radius,~$\langle r^2 \rangle$,
the magnetic moment $\mu$ and the quadrupole moment $Q_2$, which are given by \cite{Pacheco97,Cardarelli95}:
\begin{eqnarray}
\langle  r^2 \rangle = \lim_{q^2 \rightarrow 0} 
\dfrac{6 [G_0(q^2)-1] }{q^2} \ , \
      \mu = \lim_{q^2 \rightarrow 0} G_1 (q^2) \ , \
      Q_2 = \lim_{q^2 \rightarrow 0} 
      \dfrac{3 \sqrt{2} G_2 (q^2)}{q^2}~.
\end{eqnarray}

In addition we calculate the $\rho$-meson decay constant $f_\rho$ defined by the matrix element ~\cite{Zuber1980}:
\begin{eqnarray}
      \langle 0| \bar{d}\Gamma^\mu u |\rho; i \rangle \dfrac{Q_u + Q_{\bar{d}} }{\sqrt{2}} =  i 
    \sqrt{2} f_\rho m_\rho \epsilon^\mu_i \ . 
\end{eqnarray}  
where $Q_{u}=2/3$ and  $Q_{\bar{d}}=1/3$ 
are the charges of quark and antiquark. The matrix element  of the electromagnetic current between the vacuum and $\rho-$meson states is computed from 
a loop integral, and we choose the plus momentum component and the polarisation $\epsilon_z$ defined in (\ref{polar}). After integrating in $k^-$ 
the result is:
  \begin{eqnarray}
   f_\rho =  \dfrac{N_c N}{m_\rho} \int \dfrac{d^2 k_\perp}
   {(2 \pi)^3} \int_0^1 dx\dfrac{  4 \left( - x ( p^{+} + p^{+^2})+ k_{\perp}^2 
 + m^2 \right) -  
 \dfrac{m_{\rho}}{2}\dfrac{ 4m p^+(2x
  - 1)(k^-_{on} - x p^+)}{ \left[
  p^+ (k^-_{on} + x p^+)/2  + m_{\rho}m \right] }}
  {x (1 - x)^3 (m^2_\rho - M^2_0 ) (m^2_\rho-M^2_R )^2} 
  \label{cc26} \ ,
\end{eqnarray} 
where $p^{\mu}=(m_\rho,\vec{0})$, $N_c=3$, is number of colors 
and $k^-_{on}=(k^2_{\perp}+m^2)/x$ is the particle on-mass-shell relation. The other quantities in (\ref{cc26})  are:
\begin{eqnarray}
 M^2_0 = \dfrac{\vec{k}^2_{\perp}+m^2}{x} + \dfrac{(\vec{p}_{\perp}-\vec{k}_{\perp})^2+ m^2}{1 - x} - p^2_\perp \ ; \   
   M^2_{_R} = \dfrac{   \vec{k}^2_{\perp}+m^2}{x} + 
   \dfrac{(\vec{p}_{\perp}-\vec{k}_{\perp})^2+ m_R^2}{1 - x} - \vec{p}^2_\perp ~,
   \end{eqnarray}
In order to have bound state, the 
following conditions are necessary,~$m> m_{\rho}/2$~e~$m_{R}+m > m_{\rho}$,~where $m_R$ is the 
regulator mass. It is worthwhile to observe that the expression for the decay constant is free of zero-mode contributions, which are suppressed by
our particular covariant choice of the vertex function.

\begin{table}[tbh!]
\begin{center}
\begin{tabular}{|c|c|c|c|c|}  
\hline
Model & $ f_{\rho}~[GeV]$ & $\langle r^2 \rangle~[fm^2]$ & $\mu~[e/2m_{\rho}]$ & $Q_2~[e/m^2_{\rho}]$ \\ 
 \hline 
our work      &  0.154   & 0.268  & 2.21 & -0.882  \\
 \cite{Bakker2002}  &   0.134  & 0.296 & 2.10  & -0.910   \\ 
\cite{Roberts2011}  &   0.130  & 0.312 & 2.11  & -0.850    \\  
 \cite{Bhagwat2008} &   0.207  & 0.540 & 2.01  & -0.410     \\ 
\hline 
\end{tabular}
\caption{$\rho$ meson electroweak static observables. We compare our model 
with others ones from the literature~\cite{Bakker2002,Roberts2011,Bhagwat2008}. 
In our model the quark mass is 0.430 GeV and the regulator mass is $3$ GeV.
}
\end{center}
\end{table}

\section{Numerical Results}

\begin{figure}[tbh!]
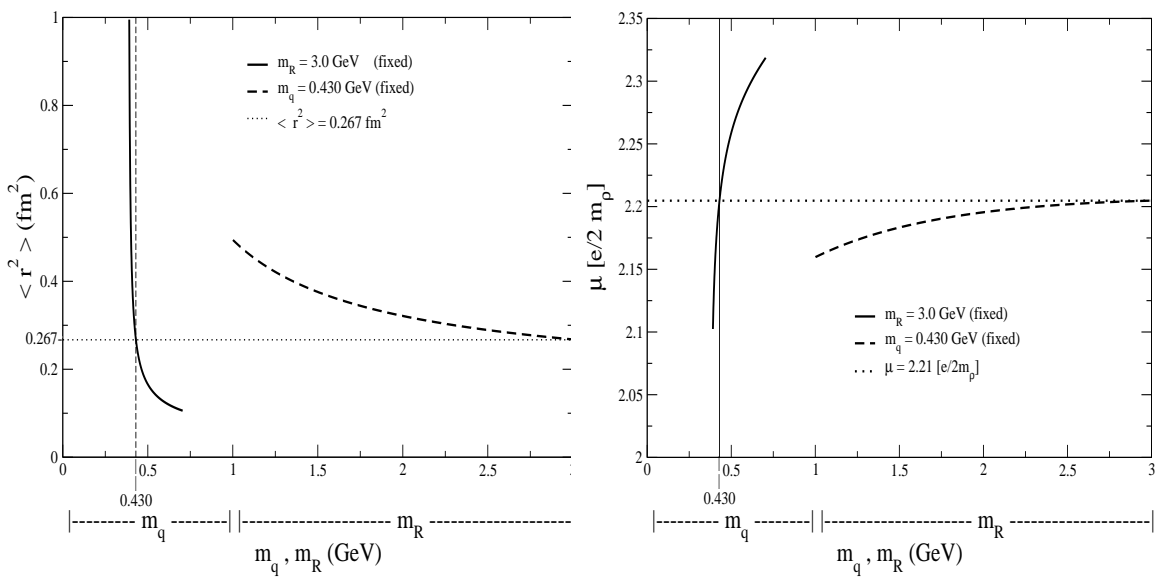

\epsfig{figure=fig21v1.eps,width=7.50cm,height=7.50cm}
\epsfig{figure=fig22v1.eps,width=7.50cm,height=7.50cm}
\begin{center}
\caption{$\rho$ meson electromagnetic radius (left frame) 
and magnetic moment (right frame). Variation with the regulator mass with fixed 
quark mass (dashed line) and variation with the quark mass with fixed regulator mass (solid line). } \label{fig1}
\end{center}
\end{figure}

\begin{figure}[tbh!] 
\begin{center}
\epsfig{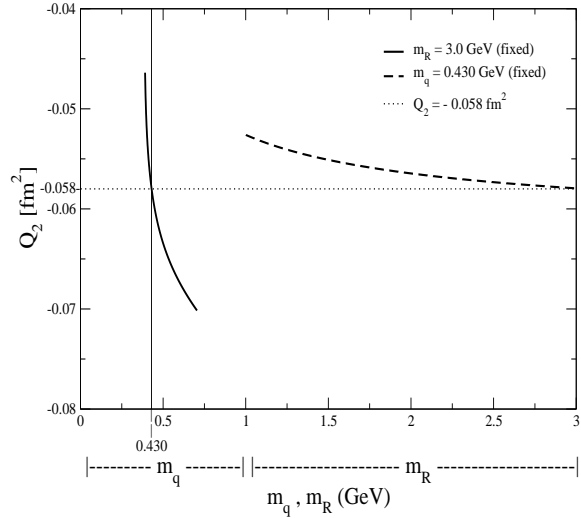}
\end{center}
 \caption{Quadrupole momentum of the $\rho$ meson. 
 Variation with the regulator mass with fixed quark mass (dashed line) and variation with the quark mass with fixed regulator mass (solid line).} \label{fig2}
\end{figure}

The input for the numerical calculations of the $\rho$ meson electroweak observables are 
the quarks mass and regulator mass.  The experimental decay 
constant~$f^{exp}_{\rho}=0.152\pm0.008$~GeV~\cite{PDG2014} is fitted
with the Eq.~(\ref{cc26}) for $m=0.430$~GeV and $m_R=3$~GeV.  In table I, 
we compare  our results with other models ~\cite{Bakker2002,Roberts2011,Bhagwat2008}. 
The qualitative trend for the static electromagnetic observables seen in the table, 
indicates that for increasing charge radius the magnetic moment decreases, while the quadrupole moment increases.  
It is suggestive that the contribution of the orbital motion to the magnetic and 
quadrupole moments is in part due to relativistic effects present in the quark spin coupling to form the composite state. 
It is natural that relativistic effects decrease as the charge radius increases and the quarks slow down. 
This general trend is further explored by computing the static observables for 
different regulator and quark masses in figures \ref{fig1} and \ref{fig2}. 

\vspace{0.5cm}

In figure \ref{fig1}, we  show the dependence of the charge radius and 
magnetic moment with the regulator and quark masses. We varied each mass 
independently while keeping the other one fixed. In the right frame of  figure~\ref{fig1}, 
we observe that the electromagnetic radius 
decreases by increasing any of the mass scales. The effect is sharp for variations 
in the quark mass, because in our model the quarks forms a 
bound state, with minimum quark mass of $m_\rho/2$, when the meson is 
dissolved in the continuum. Towards this limit the size grows to infinity and $\mu\to 2$, as clearly seen 
in the figure. The results for the quadrupole moment presented in figure \ref{fig2} corroborate the 
general trend already discussed together
with table I. As the mass scales decrease the quadrupole moment tends to have its magnitude decreased.


\section{Summary}

In the present work, we adopted the prescription proposed in
 Refs.~\cite{Inna84,Inna89} to compute the $\rho$-meson electromagnetic form factors.  The  matrix element
$I^+_{00}$  in the light-front spin basis is not used in the calculation of the form factors, and due to that the contribution of
 non-valence or zero-mode terms are not present in these observables \cite{Choi2004,Pacheco2012}. 
Using the Bethe-Salpeter amplitude model for the $\rho$ meson proposed in~\cite{Pacheco97}, 
we calculate the decay constant, charge radius,  magnetic and quadrupole moments. 
We perform a quantitative analysis of the static electromagnetic observables  with the variation of the model parameters, namely  
the constituent quark  and regulator masses. The results obtained for the charge radius in
the present model are in agreement with those expected for a composite state of nonrelativistic constituent quarks: 
the radius decreases by increasing the quark mass, the same behaviour is also found when  the regulator mass increases.
 
The values of constituent quark mass~$m=0.430$~GeV and  regulator 
mass~$m_R=3.0$~GeV give $f_{\rho}=0.154$~GeV~(see also table I), which is close 
to the experimental decay constant of $f^{exp}_{\rho}=0.152 \pm 0.008$~GeV \cite{PDG2014}.  
In this case we compare our results for the charge radius,  magnetic and quadrupole moments to the outcome of other models found in the 
literature~\cite{Bakker2002, Roberts2011, Bhagwat2008}. We note that the magnetic moment is pretty much close to 2 in all calculations, 
while a larger variation is found for the decay constant, charge  radius and quadrupole moment. The magnetic moment is more robust to model variations, as it approaches the expected sum of individual quark magnetic moments, while the other observables depends crucially on the spatial distribution of the quarks, which is more sensitive to the details of the model.


{\bf Acknowledgments.} 
This work was supported in part by the
Brazilian agencies FAPESP (Funda\c{c}\~ao de Amparo \`a Pesquisa do
Estado de S\~ao Paulo), CNPq (Conselho Nacional de
Desenvolvimento Cient\'\i fico e Tecnol\'ogico) and CAPES 
(Coordena\cc \~ao de Aperfei\cc oamento de Pessoal de 
N\'ivel Superior). Also, we thank the {\it Centro 
Nacional de Supercomputa\cc \~ao~(CESUP),
~Universidade Federal do Rio Grande do Sul} by the 
use of computacional facilities.



\begin{thebibliography}{3}

\bibitem{PDG2014}
The Review of Particle Physics
K.A. Olive et al. (Particle Data Group), Chin. Phys. C, 38, 090001 (2014). 

\bibitem{Yndurain} F.~J.~Yndur\'ain:~{\it The theory of Quarks and Gluons Interactions}, 
~Spring Verlag~(1983).

\bibitem{Smilga} Andrei~Smilga:~{\it Lectures On Quantum Mechanics}, 
~Word Scientific~(2001).

\bibitem{Close1980} F.~E.~Close:
~{\it An Introduction to Quarks and Partons}, Academic Press~(1980).

\bibitem{Roberts96} C.~D.~Roberts:~{\it 
Electromagnetic pion form-factor and neutral pion decay width.}
~Nucl.~Phys.~A{\bf A605},~475,~(1996)~475.

\bibitem{Pacheco97}
J.~P.~B.~C. de Melo and~T.~Frederico:
{\it Covariant and light front approaches to the rho meson electromagnetic form-factors.}
Phy. Rev. C~{\bf 55},~2043~(1997).

\bibitem{deMelo99} J.~P.~B.~C.~de~Melo,~H.~W.~L.~Naus and 
T.~Frederico:~{\it 
Pion electromagnetic current in the light cone formalism.}
~Phys.Rev. C {\bf 59},~2278~(1999). 


\bibitem{Choi99} H.~M.~Choi and C.~R.~Ji:~{\it Mixing angles and electromagnetic properties of 
ground state pseudoscalar and vector meson nonets in the light cone quark model}, 
Phys.~Rev.~D{\bf 59},~074015~(1999).



\bibitem{deMelo2003} 
  J.~P.~B.~C.~de Melo and T.~Frederico,
  Braz.\ J.\ Phys.\  {\bf 34}, 881 (2004)
  [hep-ph/0407304].


\bibitem{Pacheco2002}
J.~.P.~B.~C.~de~Melo, T.~Frederico, E.~Pace, and G.~Salm\'e,
Nucl.~Phys.~A{\bf ~07},~(2002)~399; 
ibid.,{\it Frame dependence of the pair contribution to the pion electromagnetic 
form-factor in a light front approach.}~Braz.~J.~Phys.~{\bf 33},~301~(2003). 


\bibitem{deMelo2004} J.~P.~B.~C.de Melo,~T.~Frederico,~
E.~Pace and G.~Salm\'e:~{\it Electromagnetic form-factor of the pion 
in the space and time - like regions within the front form dynamics.}~Phys.~Lett.~B~{\bf 581}~75~(2004).
ibid,:{\it Space-like and time-like pion electromagnetic 
form-factor and Fock state components within the light-front dynamics.}
~Phys.~Rev.~D~{\bf 73},~074013~(2006).


\bibitem{Otoniel2012} Edson~O.~da Silva,~J.~P.~B.~C.~de Melo,~
B.~El-Bennich and 
Victo~S.~Filho:~{\it Pion and kaon elastic form factors in a refined light-front model.}
~Phys.~Rev.~C{\bf 86},038202~(2012).
 
\bibitem{Brodsky98}
Stanley J.~Brodsky and Hans-Christian~Pauli and Stephen~Pinsky:{\it Quantum Chromodynamics and Other Field
Theories on the Light Cone.}~Physical Report,~{\bf 301},~(1998),~299.  

\bibitem{Harindranath2000} A~Harindranath:~{\it 
Ligh-front quantum chromodynamics:~Towards phenomenology},
~Pramana Journal of Physics,~{\bf 55},~241~(2000).

\bibitem{Beane2013}
Silas R. Beane:~{\it Broken Chiral Symmetry on a Null Plane.}
Annals Phys. {\bf 337},~111~(2013).


\bibitem{Jaus1990}
W.~Jaus:~{\it Semileptonic Decays of B and d Mesons in the 
Light Front Formalism.}~Phys.~Rev.~D{\bf 41},~3394~(1990).

\bibitem{Pacheco2009} J.~P.~B.~C.~de Melo, T.~Frederico, E.~Pace, 
S.~Pisano, G.~Salm\'e:~{\it Time- and Spacelike Nucleon Electromagnetic Form Factors 
beyond Relativistic Constituent Quark Models.}~Phys.~Lett.~B{\bf 671},~153~(2009). 
 
\bibitem{Kazuo2014} J.~P.~B.~C.~de Melo,~K.~Tsushima,~Bruno~El-Bennich,
~E.~Rojas,~T. Frederico:{\it Pion structure in the nuclear medium.}
~Phys.~Rev.~C{\bf 90},~035201~(2014).

\bibitem{Frankfurt79} L.~L.~Frankfurt and M. Strikman:
{\it Relativistic Description of the Deuteron: Elastic and Inelastic 
Form-Factors in the Region Kinematically Forbidden 
for Scattering on a Free Nucleon.}, 
Nucl. Phys. B{\bf 148}, 107 (1979), ibid,
~L.L. Frankfurt and M. Strikman:
{\it High-Energy Phenomena, Short Range Nuclear Structure and QCD.}~
Phys. Rep. {\bf 76}, 215~(1981).


\bibitem{Inna84} I.~L.~Grach and  L.~A.~Kondratyuk: 
{\it Electromagnetic Form-factor Of Deuteron In Relativistic Dynamics. 
Two Nucleon And Six Quark Components.}
~Sov. J. Nucl. Phys. {\bf 39}, 198 (1984).


\bibitem{Chung88} P.L.Chung, 
F. Coester, B. D. Keister and 
W.N.Polizou:~{\it Hamiltonian Light Front Dynamics of Elastic Electron Deuteron Scattering.}~
Phys. Rev.,~{\bf C37},~(1988),~2000.



\bibitem{Inna89} I.~L.~Grach, L.~A.~Kondratyuk, and M.Strikman:
{\it Is the structure In the Deuteron magnetic form-factor 
at $Q^2$ approximately = $2-Gev^2$ a new evidence 
for nuclear core?}.~Phys. Rev. Lett. {\bf 62}, 387 (1989).

\bibitem{Brodsky92}
S.~J.~Brodsky and J.~R.~Hiller:{\it Universal properties of the electromagnetic interactions of
spin one systems.}~Phys.~Rev.~{\bf D46},~(1992),~2141.


\bibitem{Frankfurt93} L.~L.~Frankfurt, T.~Frederico, and
M.~Strikman:
{\it Deuteron form-factors in the light cone quantum mechanics 
'good' component approach.}~Phy.~Rev.~{\bf C48}, 2182 (1993).


\bibitem{Cardarelli95} F.~Cardarelli and I.~L.~Grach and 
I.~M.~Narodetsky,~G.~Salme and S.~Simula:{\it Electromagnetic 
Form-Factors of the Rho Meson in a Light Front
Constituent Quark Model.}~
Phys.~Lett.,~{\bf B349},~(1995),~393. 



\bibitem{Ji2001} B.~L.~G.~Bakker and C.~R.~Ji:~{\it The Vector meson form-factor analysis in light front 
dynamics.}~Phy.~Rev.~D{\bf 65},~116001~(2002).


\bibitem{Bakker2001}
Bernard L.~G.~Bakker,~Ho-Meoyng Choi and Chueng-Ryong~Ji:~
{\it Regularizing the fermion loop divergencies in the light front meson currents.}
~Phys.~Rev.~D{\bf 63},~-74014~(2001).

\bibitem{Melo2004}
J.~P.~B.~C.~de Melo and T.~Frederico:~{\it Spin-1 particle in the light-front approach.}~
Braz.~J.~Phys.~{\bf 34},~881~(2004).


\bibitem{Choi2004} Ho-Meoyng Choi and C.~R.~Ji:
{\it Electromagnetic structure of the rho meson in the light 
front quark model.}~Phy.~Rev.~D{\bf 70},~053015~(2004).

\bibitem{Pacheco2012}
 J.~P.~B.~C.~de Melo, and T.~Frederico:~{\it Light-Front Projection of Spin-1 Eletromagnetic
Current and Zero-Modes.}~
Phys.~Lett.,~{\bf B708 }, (2012), 87.  
 

\bibitem{Zuber1980}{C.~Itzykzon~and~J.~B~Zuber},~{\it Quantum Field Theory.},
~New york,~(1980).

\bibitem{Bakker2002} 
  B.~L.~G.~Bakker, H.~M.~Choi and C.~R.~Ji,
  ``The Vector meson form-factor analysis in light front dynamics,''
  Phys.\ Rev.\ D {\bf 65},~116001~(2002). 

  
  
\bibitem{Roberts2011} 
  H.~L.~L.~Roberts, A.~Bashir, L.~X.~Gutierrez-Guerrero, C.~D.~Roberts and D.~J.~Wilson,
  ``pi- and rho-mesons, and their diquark partners, from a contact interaction,''
  Phys.\ Rev.\ C {\bf 83}, 065206 (2011).

  
  
\bibitem{Bhagwat2008} 
  M.~S.~Bhagwat and P.~Maris,
  ``Vector meson form factors and their quark-mass dependence,''
  Phys.\ Rev.\ C {\bf 77}, 025203 (2008).
 
\end{thebibliography}
  \end{document}